\documentclass[11pt,a4paper]{article}
\pdfoutput=1

\usepackage{subfig}
\usepackage{amsmath, amssymb, amsthm} 
\usepackage{graphicx}

 \usepackage{hyperref}
 \hypersetup{
    colorlinks,%
    citecolor=black,%
    filecolor=black,%
    linkcolor=black,%
    urlcolor=black
}

\newcommand{\MET} {\ensuremath{E_{T}^{\textnormal{miss}}}}

\begin{document}

\title{Stochastic mass-reconstruction: a new technique to reconstruct resonance masses of heavy particles decaying into tau lepton pairs}
\date{\today}
\author{
        Sho Maruyama\\ 
Fermi National Accelerator Laboratory, Batavia, IL 60510, USA\\
email: \href{mailto:mayaruma@fnal.gov}{mayaruma@fnal.gov} 
 }
\maketitle

\begin{abstract}
The invariant mass of tau lepton pairs turns out to be smaller than the resonant mass of their mother particle and the invariant mass distribution is stretched wider than the width of the resonant mass as significant fraction of tau lepton momenta are carried away by neutrinos escaping undetected at collider experiments. 
This paper describes a new approach to reconstruct resonant masses of  heavy particles decaying to tau leptons at such experiments.
A typical example is a $Z$ or Higgs boson decaying to a tau pair.
Although the new technique can be used for each tau lepton separately, I combine two tau leptons to improve mass resolution by requiring the two tau leptons are lined up in a transverse plane.
The method is simple to implement and complementary to the collinear approximation technique that works well when tau leptons are not lined up in a transverse plane.
The reconstructed mass can be used as another variable in analyses that already use a visible tau pair mass and missing transverse momentum as these variables are not explicitly used in the stochastic mass-reconstruction to select signal-like events.
\end{abstract}

\section{Introduction}
\label{sec:intro}
After the discovery of the Standard Model (SM) Higgs boson at the CERN LHC~\cite{atlasDiscH, cmsDiscH}, precision measurements of the scalar particle properties have become an important part of the physics program at collider experiments. 
Despite its larger coupling to Higgs bosons, the $h \rightarrow \tau \tau$ mode poses experimental challenges; it suffers from  lower identification efficiencies and higher mis-identification rates with respect to those of electrons and muons if tau leptons decay hadronically; or it suffers from a decrease in signal by the tau lepton branching fractions to light-flavor leptons.
Furthermore, both hadronic and leptonic modes suffer from softer and wider tau pair mass distributions due to neutrinos taking away some fraction of tau lepton momenta. 

In this paper I develop a new technique to define a new variable, stochastically reconstructed mass, without relying on a missing transverse energy ($ie$ indirect measure of neutrino momentum).
By design this technique is different from other techniques such as the collinear approximation~\cite{ca} and missing mass calculator~\cite{mmc}.
The new technique can be used for the $Z$ boson production, which provides a useful way to calibrate tau reconstruction algorithms~\cite{cmsTau}, and other heavy particles, such as a SM Higgs boson as well as a hypothetical particle like a $Z'$ particle.

The mathematical expression for momentum distributions of  $n$-body decay is derived in sec.~\ref{sec:nbody}, followed by development of a new mass reconstruction technique in sec.~\ref{sec:smr}. 
Studies of $Z$, $h$, $Z'$ boson, and QCD di-jet productions are presented in sec.~\ref{sec:z2tautau},~\ref{sec:h2tautau}, and~\ref{sec:qcd}.
Section~\ref{sec:note} contains technical notes on biases induced by kinematic selection, and sec.~\ref{sec:summary} summarizes the paper.

\section{Unconstrained $n$-body decay}
\label{sec:nbody}
First, I explicitly describe momentum distribution for small $n$ up to 4, and then generalize the obtained results.
I assume spin correlation among mother and daughter particles is negligible as well as mother particle mass is much greater than daughter particle masses, so that the masses of daughters are negligible. 
Validity of the assumptions is discussed in sec.~\ref{sec:z2tautau}.
The simplest case is a 2-body decay and there is no degree of freedom left in this case; $ie$, once momentum of one out of  two particles is randomly chosen, then the other particle has a momentum of $C(1-x)$, where $C$ is the momentum carried by the mother particle and $x$ is the fraction of momentum carried by the first daughter.
For the sake of simplicity, I set $C$ to unity hereafter.

The result for 2-body decay  means that momentum distribution is flat $(ie$, all values of possible momentum are equally likely).
The result can be expressed as 
\begin{eqnarray}
f(x)_{2\text{-body}} = 1 \times (1-x)^{0} = 1.
\end{eqnarray}

Now let us study a 3-body decay case.
As in the 2-body decay, I can pick some fraction of momentum for the first particle out of three.
And then I assign some fraction to the second particle from $(1-x)$. 
Once momenta are assigned to the first two particles, then the third one receives the left-over momentum.
This means that the momentum distribution is proportional to $(1-x)$.
By integrating it out to fix the normalization, I obtain the distribution as
\begin{eqnarray}
f(x)_{3\text{-body}} = 2 (1-x)^{1}.
\end{eqnarray}

For illustrative purposes, I take a 4-body decay case as the last explicit example.
The momentum distribution is given by 
\begin{align}
f(x)_{4\text{-body}} & \propto \int_{0}^{1-x} (1-x-y)dy \\
& =  (y\times(1-x) - 1/2 \times y^2) |_{0}^{1-x} \\
& =  ( (1-x)^2 - 1/2 \times (1-x)^2) \\
f(x)_{4\text{-body}} & = 3 (1-x)^2.
\end{align}
The last step is completed by integrating the probability to unity. 

I can continue on deriving an expression for $n$-body decay, and the result can be concisely summarized by the beta function:
\begin{eqnarray}
f(x)_{n\text{-body}} = (n-1) (1-x)^{n-2}.
\end{eqnarray}

The $(1-x)$ term is derived by integrating out the degrees of freedom left for momentum assignments.
The degrees of freedom for the $(n-1)$-th particle is always $(1- \sum_{i=1}^{n-2}{x_{i}})$ where $x_{i}$ is the momentum fraction assigned to the $i$-th daughter for $n > 2$.
This function is integrated from zero to $(1- \sum_{i=1}^{n-3}{x_{i}})$ with respect to $x_{n-1}$ for $n > 3$. 
The integration with respect to the next $x_{i}$ is repeated until all of them are integrated out and expressed in terms of $x$.
There are $(n-2)$ integrations involved because the last daughter automatically receives the left-over momentum, and the degrees of freedom associated with the first daughter is represented by $x$. 
Therefore an expression for the $n$-body momentum distribution is proportional to $(1-x)^{n-2}$.
The normalization factor is derived naturally from the power of $(1-x)$ term.

\subsection{Expectation value of $n$-body decay momentum distribution}
\label{sec:expectation}
The relevant variable to the stochastic mass-reconstruction technique is the mean value of daughter momenta in the $n$-body decay.
The mean is given by 
\begin{eqnarray}
\langle x_{n\text{-body}}\rangle = \int_{0}^{1} xf(x)_{n\text{-body}}dx = \int_{0}^{1} (n-1)  x(1-x)^{n-2}dx.
\end{eqnarray}
The integration is straightforward by exploiting the property of the beta function:
\begin{align}
\langle x_{n\text{-body}}\rangle &= (n-1) (1)! (n-2)! / (n)! \\
&= \frac{(n-1) (n-2)!}{ n (n-1) (n-2)!} \\
&= 1/n.
\end{align}

\subsection{Discrete case}
\label{sec:discrete}
In practice we have a finite chance of sampling daughter momenta in each event.
The mean is calculated as 
\begin{eqnarray}
\langle x\rangle = 1/m \times \sum_{i=1}^{m} x_{i}
\end{eqnarray}
where $m$ is a number of visible daughters and $x_{i}$ is momentum of $i$-th visible daughter.
For $n$-body decay, the mother momentum is estimated by
\begin{eqnarray}
\langle x_{\text{mom}}\rangle = n \times 1/m \times \sum_{i=1}^{m} x_{i}
\end{eqnarray}
where $n > m$ by assuming some of daughters are invisible.
For large $m$ with few invisible particles, 
\begin{eqnarray}
\langle x_{\text{mom}}\rangle = n/m \times \sum_{i=1}^{m} x_{i}\rightarrow \sum_{i=1}^{m} x_{i} \approx 1.
\end{eqnarray}
A sanity check can be performed for $m=n$; $ie$ all daughters in simulated samples.
\begin{eqnarray}
\langle x_{\text{mom}}\rangle = n/n \times \sum_{i=1}^{n} x_{i}= \sum_{i=1}^{n} x_{i} = 1 = x_{\text{mom}}.
\end{eqnarray}

\subsection{Application to tau leptons}
\label{sec:app}
Tau leptons decay to hadron(s) and a tau neutrino, or a lepton (electron or muon) and neutrinos, about two thirds and one third of the time, respectively~\cite{pdg}. 
For lepton and pion decay modes, momentum distribution of decay products (tau daughter) can be described by a simple beta function by treating tau daughters as massless and ignoring spin correlation among them. 

Momentum distribution of tau daughters is given by 
\begin{eqnarray}
f(x) = C(n-1) (1-x)^{n-2}
\end{eqnarray}
where $C$ is the momentum carried by the mother particle (tau lepton), and $n$ is the number of daughters in $n$-body decay modes. 
Expectation value of the daughter momentum is given by
\begin{eqnarray}
\langle P_{\text{daughter}}\rangle = C/n
\end{eqnarray}
by integrating the beta function and normalizing total probability to unity.
By inverting the relation we have the expectation value of the mother momentum
\begin{eqnarray}
\langle P_{\text{mom}}\rangle = n \langle P_{\text{daughter}} \rangle = C.
\end{eqnarray}
In other words the mother momentum can be estimated, on average, from the mean value of daughter momenta as long as $n$ is known, even if some of the daughter particles are invisible.

\section{Stochastic mass-reconstruction}
\label{sec:smr}
The concept (stochastic momentum-reconstruction) described in the previous section is applicable to each tau lepton.
However, the mean value is not useful when the number of daughters is small ($eg$, $\tau^{\pm} \rightarrow \pi^{\pm} \nu_{\tau}$ where only one particle is visible).
To improve resolution of reconstructed mother momentum, I combine daughters of two tau leptons to form a mass of hypothetical particle $X (\rightarrow \tau\tau)$, and hence the name stochastic mass-reconstruction.

The method described in this paper relies on the fact that some of tau lepton decay modes can be reconstructed at collider experiments.
Leptonic decay modes are commonly used as in refs.~\cite{atlasHiggs, cmsHiggs}, and hadronic modes can be reconstructed correctly $\approx$ 80\% of the time at the CMS experiment~\cite{cmsTau}.
In this study, I restrict my interests to lepton and pion (or kaon) modes, where charged pions and kaons are reconstructed by combining momentum and energy measurements, and neutral pions are reconstructed by combining photons as a $\pi^{0}$ decays to $\gamma\gamma$ predominantly because of the axial anomaly~\cite{bell, adler}. 
I treat charged kaons as charged pions by following the convention used in ref.~\cite{cmsTau}. 

Here I outline how the reconstruction can be performed at collider experiments.
\begin{itemize}
\item first, run tau identification algorithm to select events containing two tau leptons (or electron or muon).
\item second, keep the events if $\Delta \phi \approx \pi$ where $\Delta \phi$ is the angle between the two leptons in a transverse plane.
\item third, Lorentz transform the visible part of tau decay product along a beam direction ($ie \, z$-direction) by $-\beta_{z}^{vis}$ that is the $z$ component of the visible Lorentz boost vector times minus one. Here I work in a cylindrical coordinate system with the $z$-direction along colliding beams.
\item fourth, compute the mean of daughter particle momenta with a weight of $n$ that is the number of visible daughters plus one or two for hadronic and leptonic mode, respectively.
\end{itemize}
On the third step, I explicitly assume that the direction of visible daughters is close to the direction of its mother particle. 
This assumption is valid as long as $m_{\tau} / M_{\text{resonance}} \ll 1$ and $m_{\text{daughter}} / m_{\tau} \ll 1$. 
This fact forms the basis of the collinear approximation technique~\cite{ca}. 

In the procedure outlined above, it is noteworthy that a missing transverse energy $\MET$ plays no role.
No explicit dependence on $\MET$ signifies the conceptual difference between the stochastic mass-reconstruction technique and other methods trying to incorporate $\MET$ with visible tau momenta.
Instead the new technique exploits approximate distribution invariance controlled by a single number: daughter multiplicity.
Certain kinematic variables are known to possess similar approximate shape invariance ($eg$, angular variables in cascade decay chains of Supersymmetric particles in ref.~\cite{superRazor}).
The Lorentz boost on the third step allows us to make a connection between the decay products of two tau leptons. 
How each tau lepton decays is independent of the other tau lepton decay. 
Therefore the two decay modes are unrelated otherwise.

As implementation of tau identification algorithms varies among different colldier experiments, I perform a simple study with jets and generator-level information using the $\textsc{PYTHIA8}$~\cite{pythia} program for $Z \rightarrow \tau\tau$, $h \rightarrow \tau\tau$, and $Z' \rightarrow \tau\tau$ processes, to demonstrate how this new method works, followed by a simple study of QCD di-jet background process.
About a half of $p p \rightarrow Z \rightarrow \tau \tau$ events satisfy the criteria outlined above.
The technique is not limited to these massive bosons but applicable for any heavy particle decaying to a tau lepton pair as long as $m_{\tau} / M_{\text{resonance}} \ll 1$.
This fact is another distinct feature of the technique; $ie$ no need to prepare templates a priori by assuming specific heavy particles as long as $n$ is not small or selection bias is small ($ie$ mother particles provide large enough momentum to their daughters).

\subsection{\texorpdfstring{$Z \rightarrow \tau\tau$}{Z to tau tau}}
\label{sec:z2tautau}
To illustrate the technique with a concrete example, I take the $Z \rightarrow \tau\tau$ process generated at $\sqrt{s} $ = 13 TeV, which is the highest center-of-mass energy at the LHC.
I generate 100,000 events where the $Z$ boson mass lies between 60 and 120 GeV.
Modern tau identification algorithms utilize multi-variate analysis techniques such as artificial neural networks and boosted decision trees, which cannot be readily reproduced without accessing the software maintained by each experiment. 
To emulate tau identification algorithms, I employ the following simple selection.
\begin{itemize}
\item run the slow-jet~\cite{pythia} program, using $\textnormal{anti-}k_{t}$ clustering algorithm~\cite{antikt} with distance parameter of 0.2.
\item keep jets with $p_{T} > $ 10 GeV and pseudorapidity $\vert \eta \vert < 5.0$.
\item match jet constituents to generator-level tau daughters if jet constituents have energy $>$ 1 GeV.
\item keep events if a pair of jets has all jet constituents matched to tau daughters.
\end{itemize}
Parameter choices are motivated by tau reconstruction algorithms employed at the ATLAS experiment~\cite{atlasTau}.
Here I am not concerned with isolation requirements on tau candidates but the fact that tau daughters tend to be contained within narrow jets.
This selection affects shapes of stochastically reconstructed masses and the selection bias is discussed in sec.~\ref{sec:note}.
Light-flavor leptons are required to have energy $>$ 1 GeV but not matched to jet constituents.

\begin{figure}[!htbp]
    \centering
      \subfloat[]{
        \includegraphics[width=0.47\textwidth]{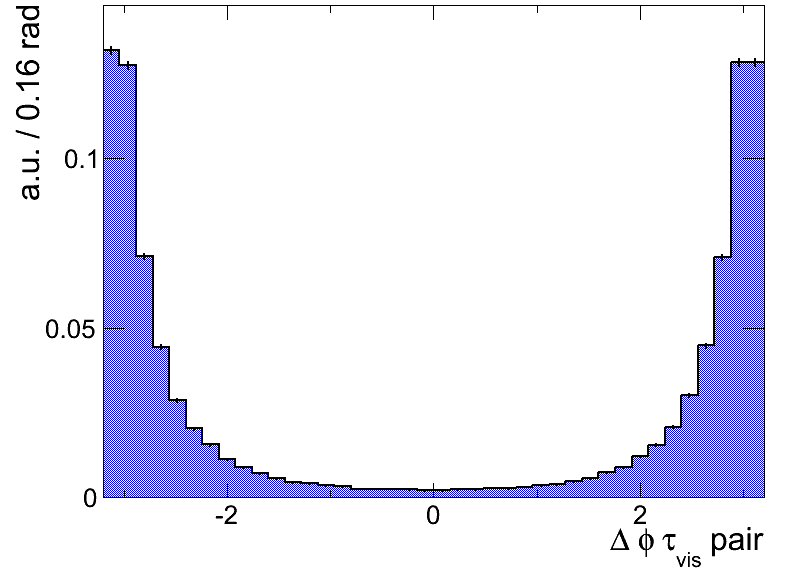}
 \label{fig:dphivis}
      }
      \subfloat[]{
        \includegraphics[width=0.47\textwidth]{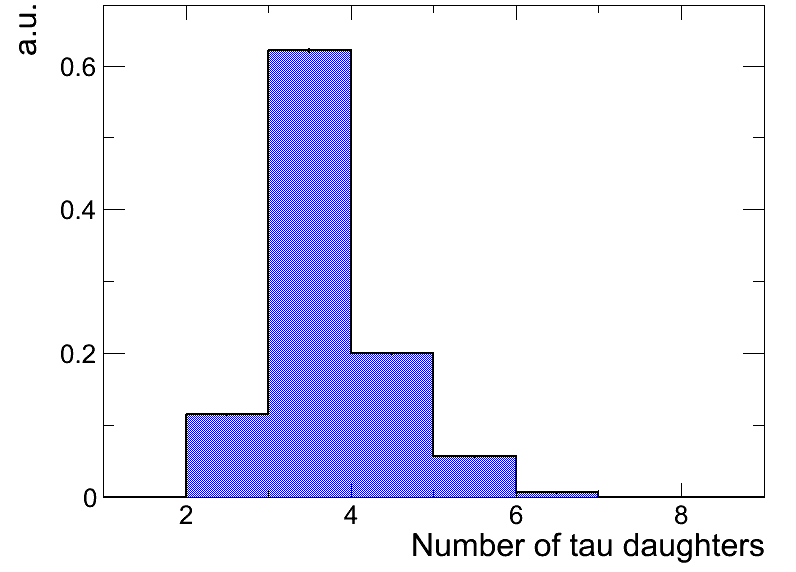}
 \label{fig:cdau1}
      }
\caption{The $\Delta \phi$ between visible tau momentum vectors, and number of tau daughters distributions in the $Z \rightarrow \tau \tau$ process.}
\end{figure}

\begin{figure}[!htbp]
    \centering
      \subfloat[]{
        \includegraphics[width=0.47\textwidth]{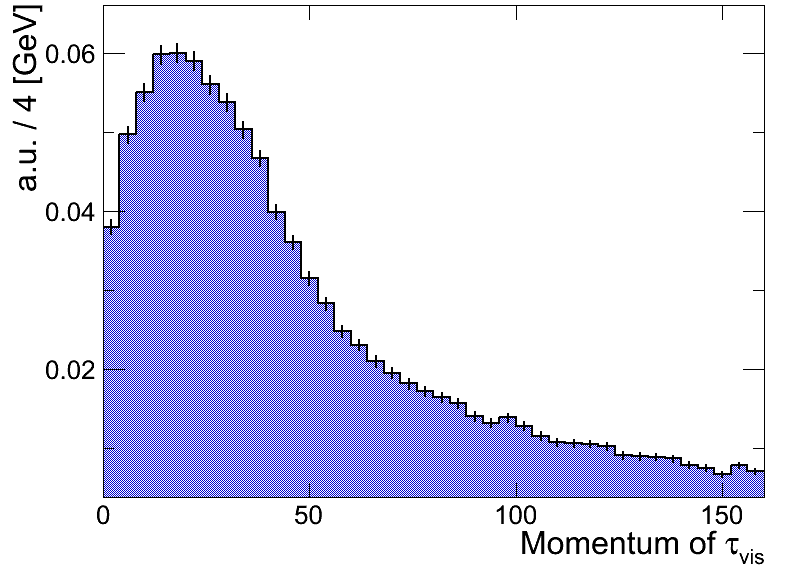}
\label{fig:pvis1}
      }
      \subfloat[]{
        \includegraphics[width=0.47\textwidth]{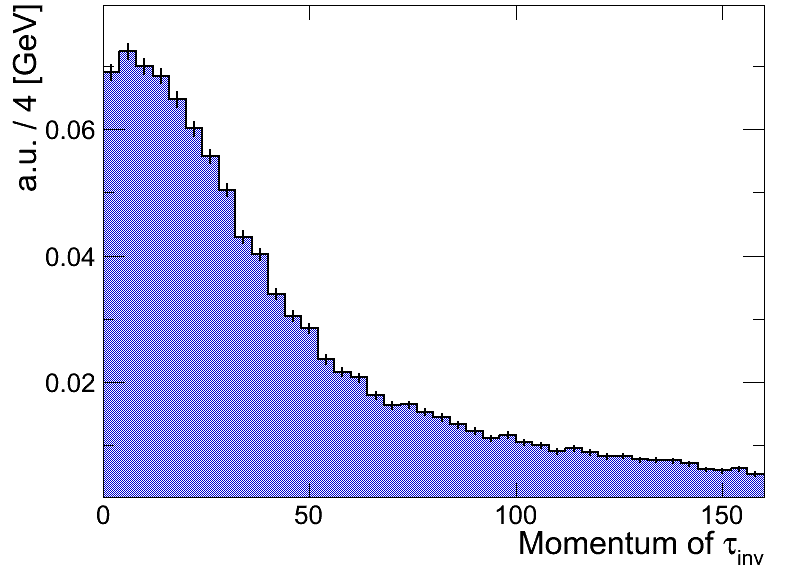}
\label{fig:pinv1}
      }
\caption{Tau visible and invisible momentum distributions in the $Z \rightarrow \tau \tau$ process.}
\end{figure}

A $\MET$ vector is defined as transverse components of the sum of all invisible particle (neutrino) momenta in an event. 
I define the visible part of tau lepton momenta as the sum of visible tau daughter 4-momenta, and the invisible part as the sum of tau daughter neutrino 4-momenta.
Figure~\ref{fig:dphivis} shows the distribution of $\Delta \phi$ between two visible tau momentum vectors. 
I select events with $\Delta \phi > \pi \times 0.9$ hereafter.
Figure~\ref{fig:cdau1} shows the number of tau daughters, including leptonic and hadronic modes.
The momentum distribution of visible and invisible tau daughters are plotted in figs.~\ref{fig:pvis1} and~\ref{fig:pinv1}.

\begin{figure}[!htbp]
    \centering
      \subfloat[]{
        \includegraphics[width=0.47\textwidth]{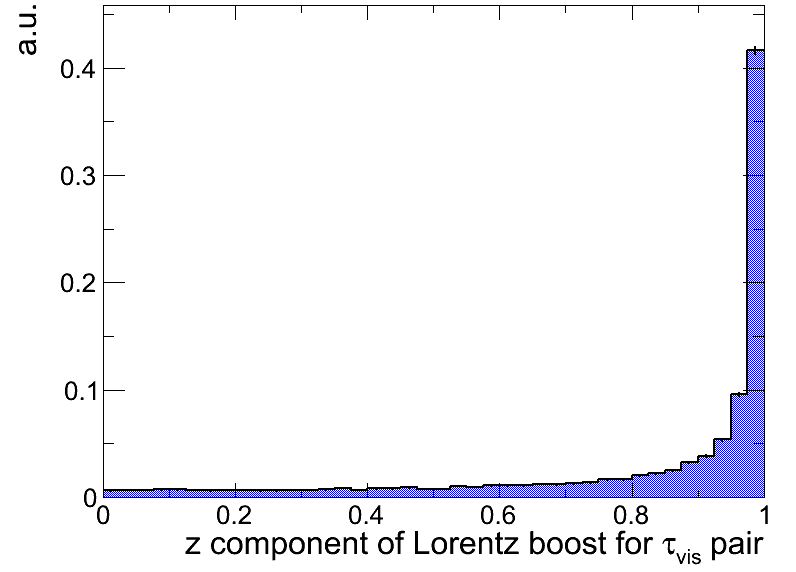}
      }
\caption{The $z$-component of Lorentz boost vector for visible tau pair in the $Z \rightarrow \tau \tau$ process.} 
\label{fig:bvis}
\end{figure}

\begin{figure}[!htbp]
    \centering
      \subfloat[]{
        \includegraphics[width=0.45\textwidth]{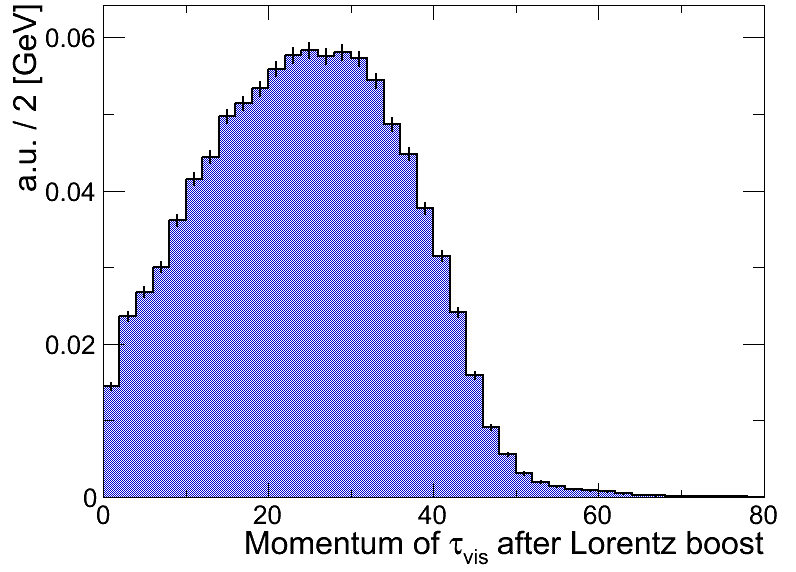}
\label{fig:pvisc1}
      }
      \subfloat[]{
        \includegraphics[width=0.47\textwidth]{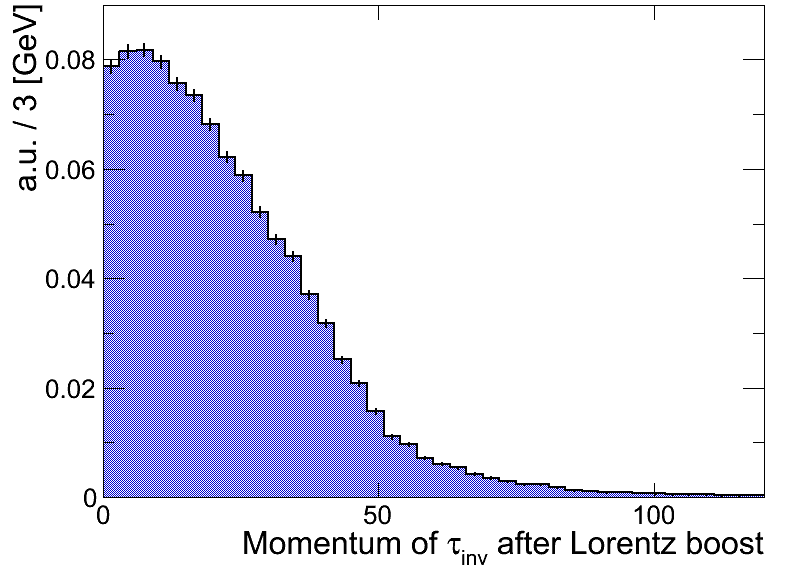}
 \label{fig:pinvc1}
      }
\caption{Tau visible and invisible momentum distributions after Lorentz boost in the $Z \rightarrow \tau \tau$ process.}
\end{figure}

Visible momenta are Lorentz transformed based on the $z$-component of visible tau pair 4-momentum (see fig.~\ref{fig:bvis}).
The transformed visible and invisible momenta are plotted in figs.~\ref{fig:pvisc1} and~\ref{fig:pinvc1}.
The procedure uses only the visible part to compute a mass.

\begin{figure}[!htbp]
    \centering
      \subfloat[]{
        \includegraphics[width=0.45\textwidth]{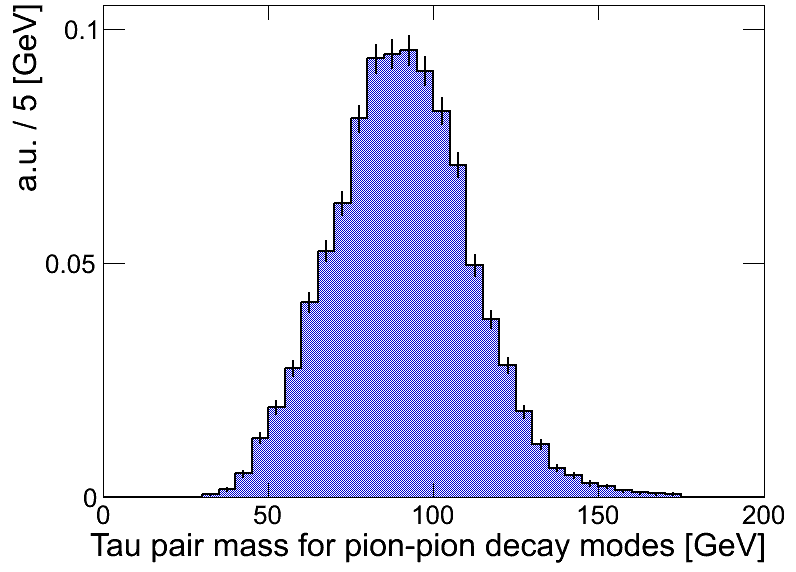}
\label{fig:msmr_pp}
      }
      \subfloat[]{
        \includegraphics[width=0.45\textwidth]{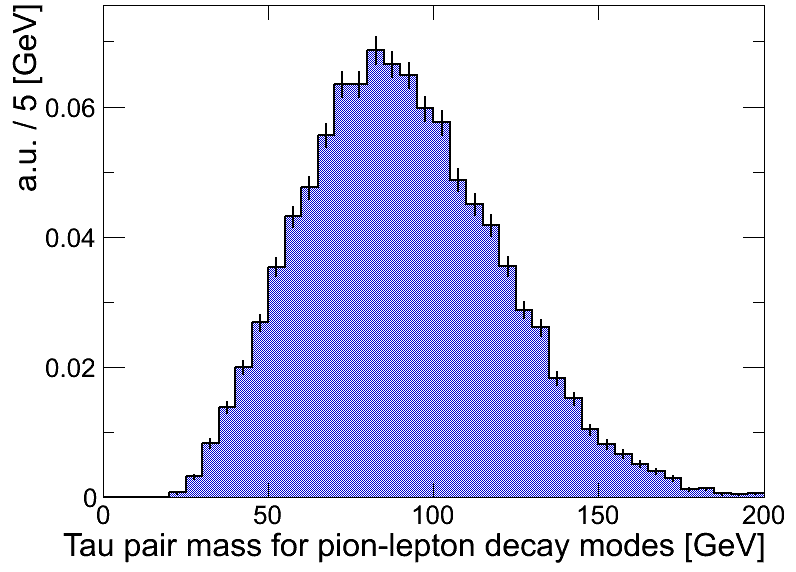}
\label{fig:msmr_pl}
      }
\hfill
      \subfloat[]{
        \includegraphics[width=0.45\textwidth]{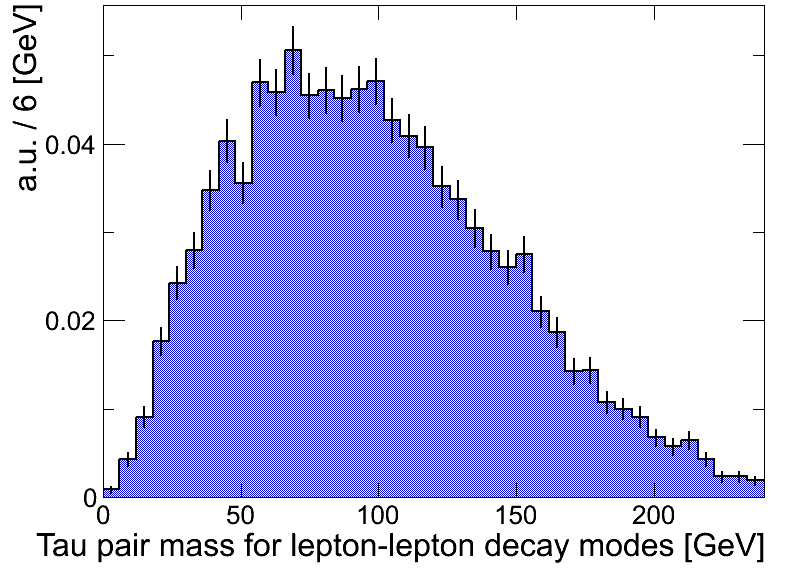}
\label{fig:msmr_ll}
      }
      \subfloat[]{
        \includegraphics[width=0.47\textwidth]{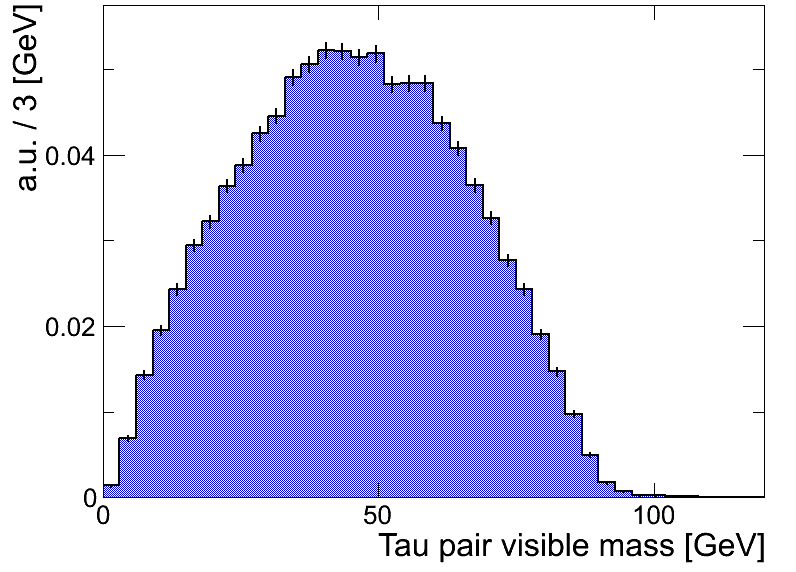}
 \label{fig:mzvis}
      }
\caption{The stochastically reconstructed mass  and visible tau pair mass distributions in the $Z \rightarrow \tau \tau$ process.}
\label{fig:msmr}
\end{figure}

\begin{figure}[!htbp]
    \centering
      \subfloat[]{
        \includegraphics[width=0.45\textwidth]{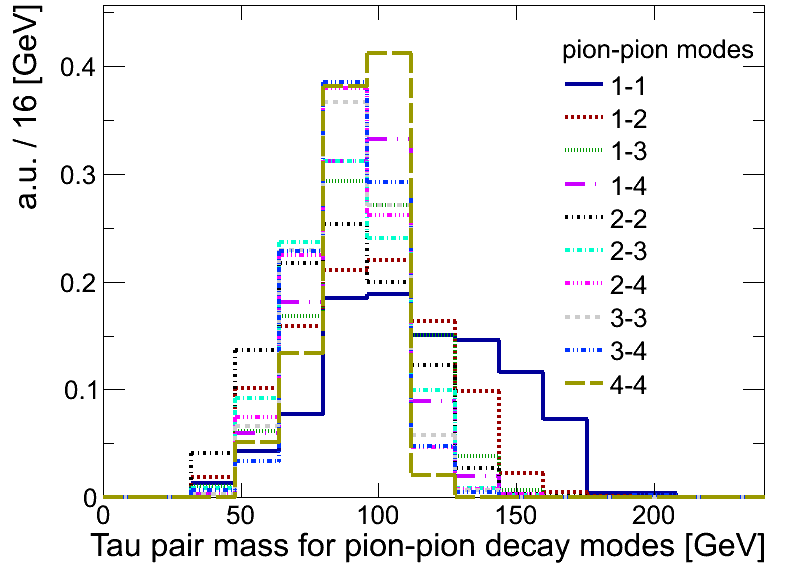}
\label{fig:msmr_pp_all}
      }
      \subfloat[]{
        \includegraphics[width=0.45\textwidth]{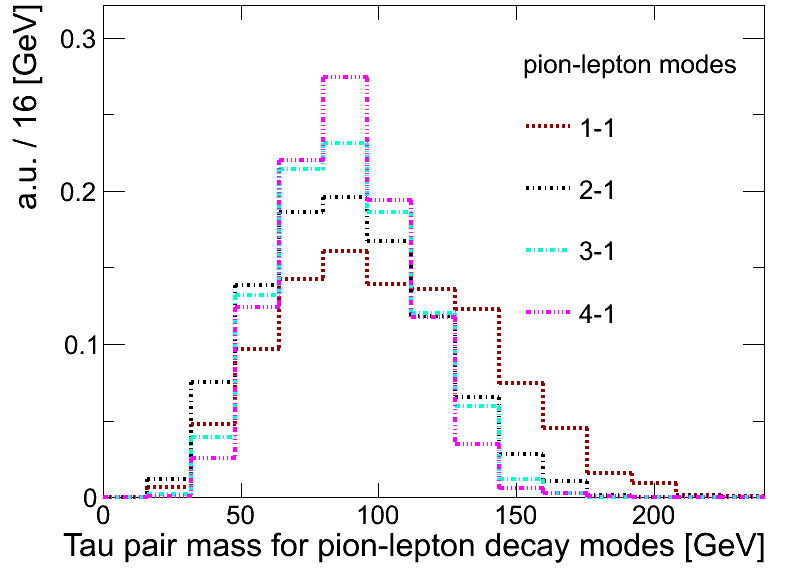}
\label{fig:msmr_pl_all}
      }
\caption{The stochastically reconstructed mass  for different combination of pion multiplicities in the $Z \rightarrow \tau \tau$ process.}
\label{fig:msmr_npion}
\end{figure}

Figure~\ref{fig:msmr} shows the stochastically reconstructed mass peaking at the $Z$ boson mass,  and visible tau pair mass.
The fact that the mass peaks at the right place lends credibility to the treatment; the momentum distributions of unconstrained $n$-body decay reasonably well describe tau lepton decays originating from heavy particles.
If this assumption were severely violated by spin correlation or kinetic constraints from massive tau daughters, the peak structure would not appear at the pole mass.
Breakdown of pion-pion and pion-lepton modes in combination of pion multiplicities is shown in fig.~\ref{fig:msmr_npion}.
The mass resolution improves for  larger pion multiplicities as expected.

\begin{figure}[!htbp]
    \centering
      \subfloat[]{
        \includegraphics[width=0.45\textwidth]{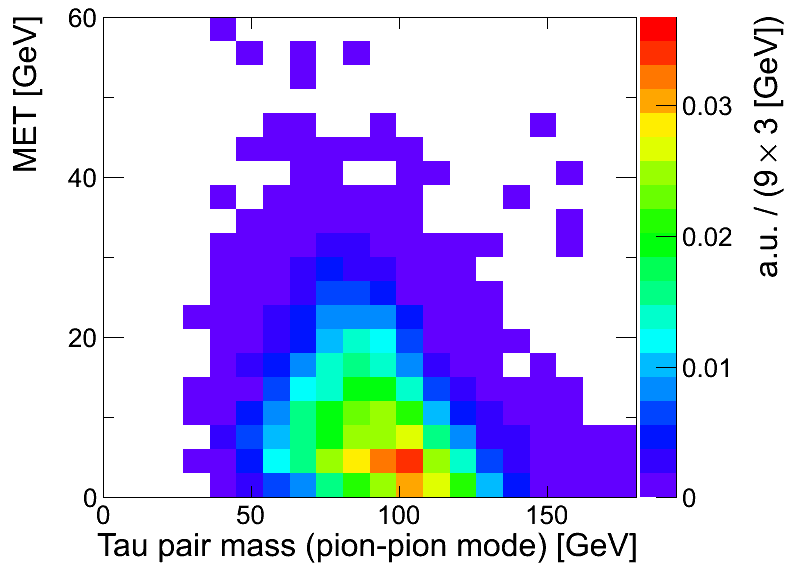}
\label{fig:msmr_pp_met}
      }
      \subfloat[]{
        \includegraphics[width=0.45\textwidth]{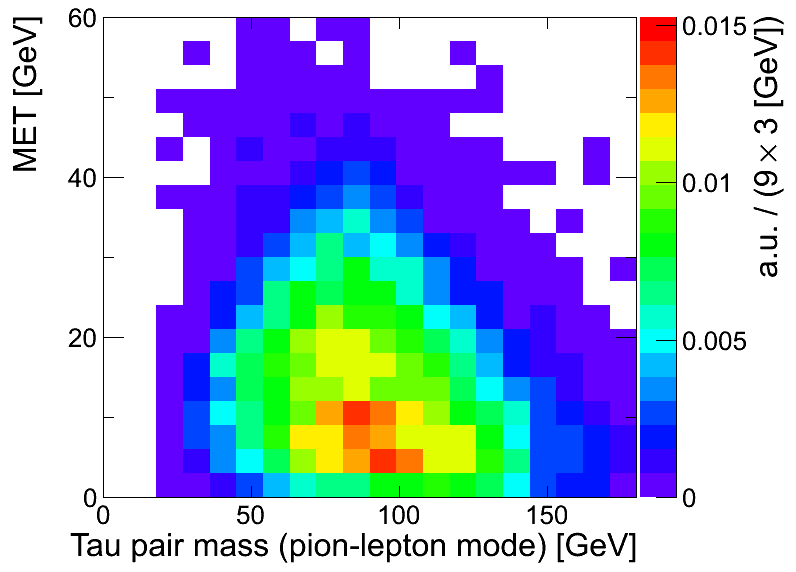}
\label{fig:msmr_pl_met}
      }
\hfill
      \subfloat[]{
        \includegraphics[width=0.45\textwidth]{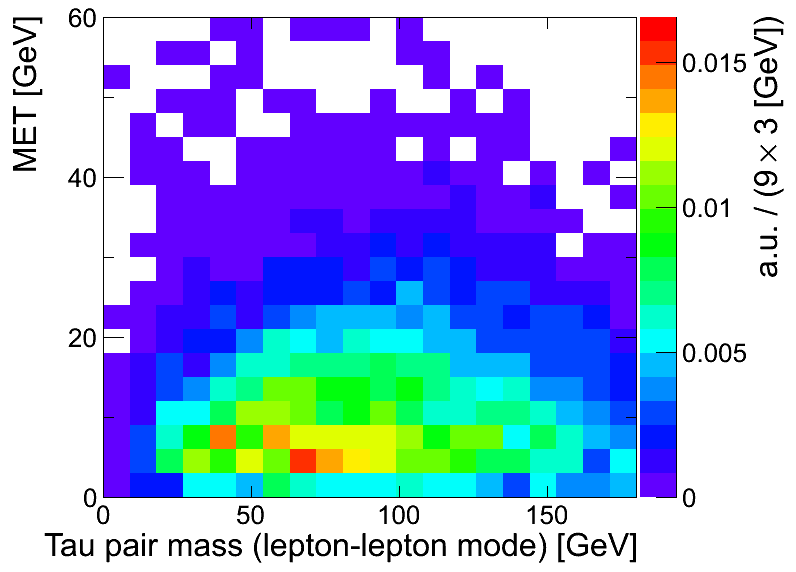}
\label{fig:msmr_ll_met}
      }
\caption{Correlation between stochastically reconstructed mass and $\MET$ in the $Z \rightarrow \tau \tau$ process.}
\label{fig:msmr_pp_2d}
\end{figure}

There is no strong linear correlation between the stochastically reconstructed mass and $\MET$  (see fig.~\ref{fig:msmr_pp_2d}).

\subsection{\texorpdfstring{$h \rightarrow \tau\tau$ and $Z' \rightarrow \tau\tau$ }{h to tau tau and Z' to tau tau}}
\label{sec:h2tautau}
\begin{figure}[!htbp]
    \centering
      \subfloat[]{
        \includegraphics[width=0.45\textwidth]{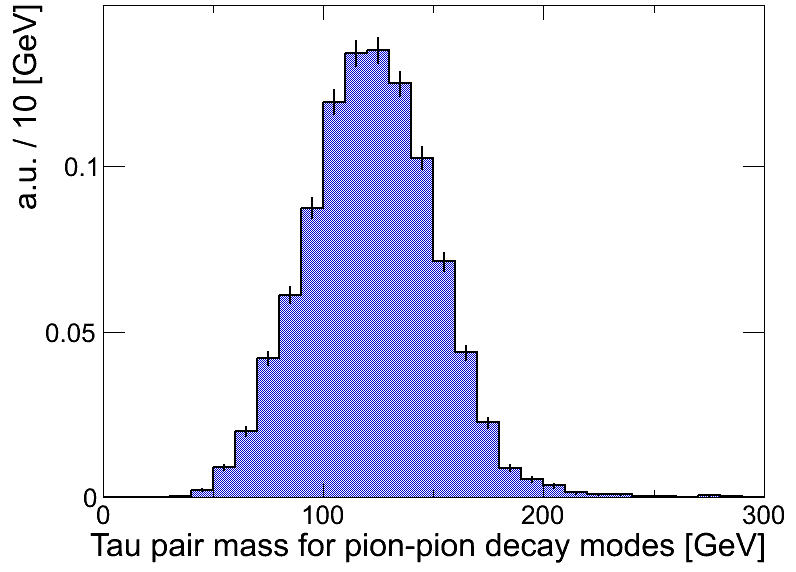}
\label{fig:msmr_pp_higgs}
      }
      \subfloat[]{
        \includegraphics[width=0.45\textwidth]{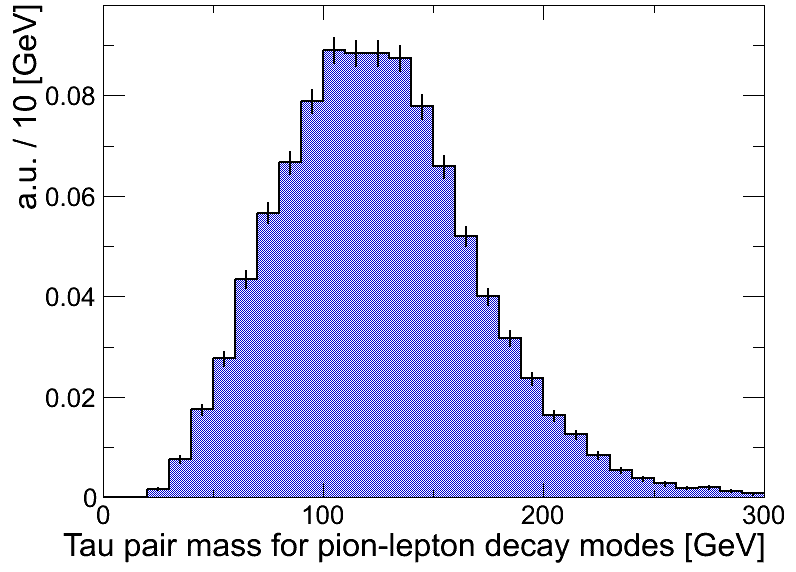}
\label{fig:msmr_pl_higgs}
      }
\hfill
      \subfloat[]{
        \includegraphics[width=0.45\textwidth]{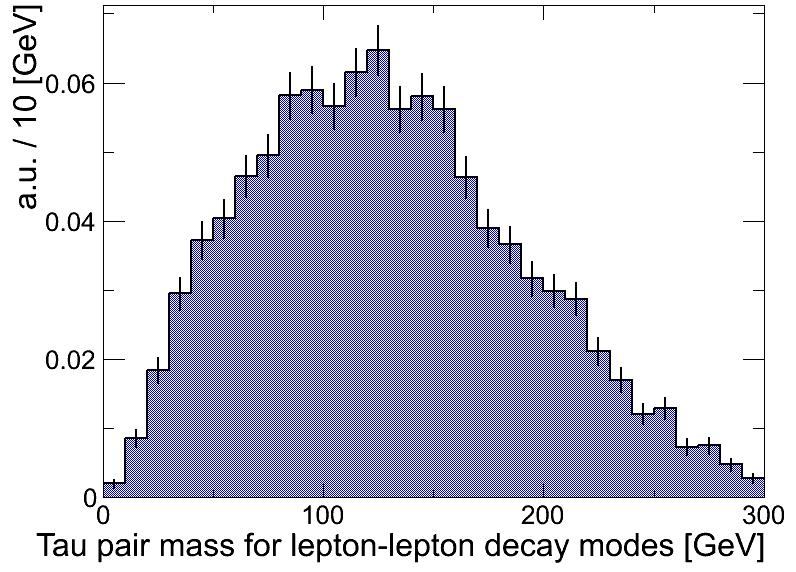}
\label{fig:msmr_ll_higgs}
      }
      \subfloat[]{
        \includegraphics[width=0.47\textwidth]{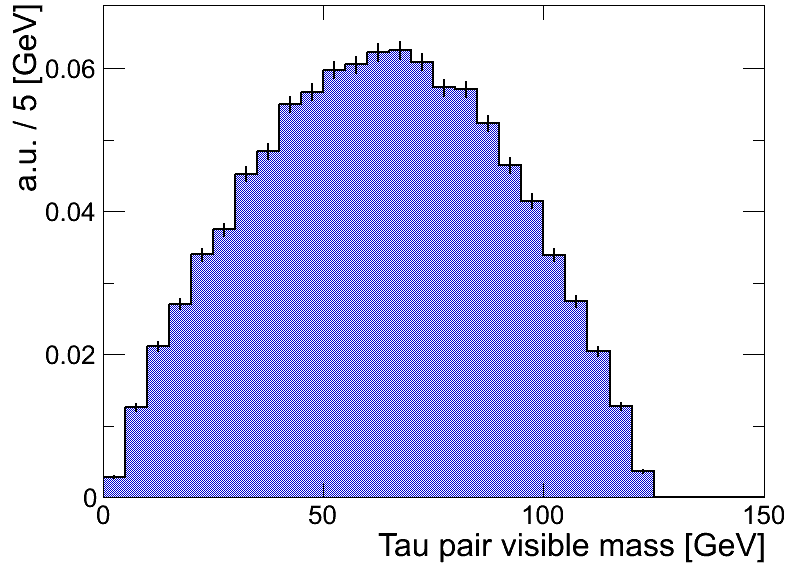}
 \label{fig:mzvis_higgs}
      }
\caption{The stochastically reconstructed mass  and visible tau pair mass distributions in the $h \rightarrow \tau \tau$ process.}
\label{fig:msmr_higgs}
\end{figure}

\begin{figure}[!htbp]
    \centering
      \subfloat[]{
        \includegraphics[width=0.45\textwidth]{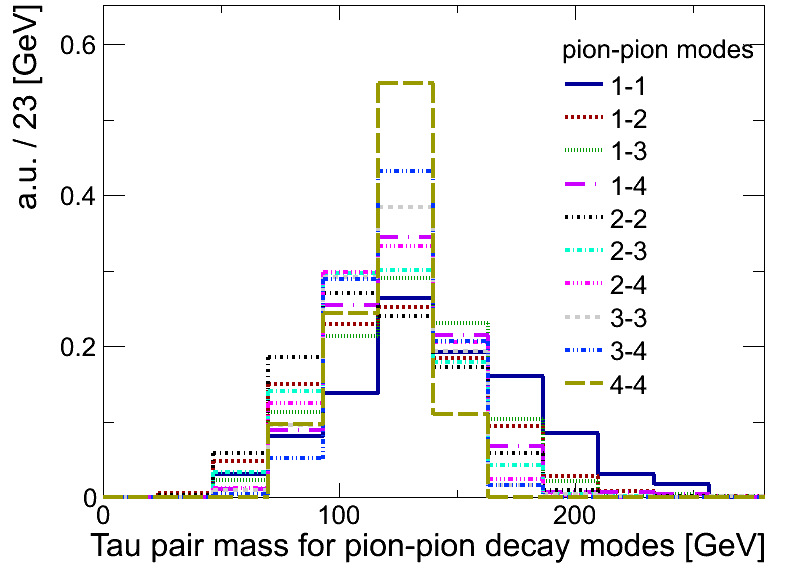}
\label{fig:msmr_pp_all_higgs}
      }
      \subfloat[]{
        \includegraphics[width=0.45\textwidth]{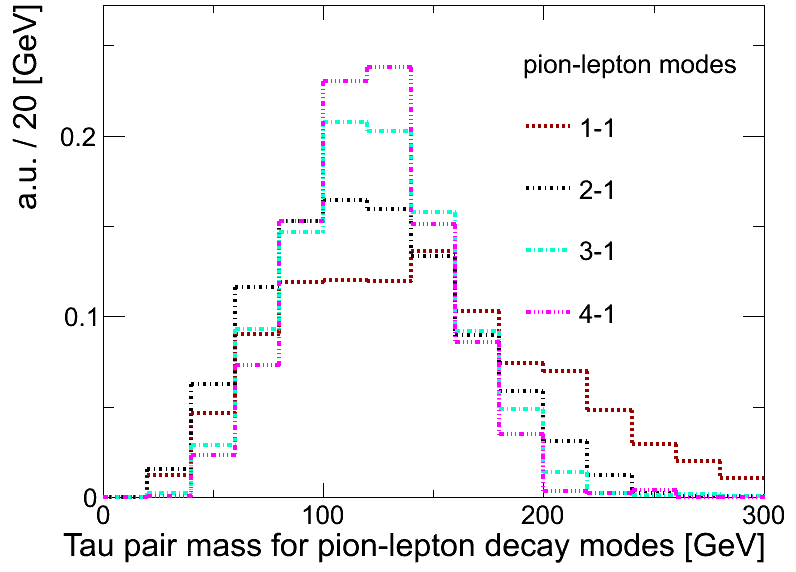}
\label{fig:msmr_pl_all_higgs}
      }
\caption{The stochastically reconstructed mass  for different combination of pion multiplicities in the $h \rightarrow \tau \tau$ process.}
\label{fig:msmr_npion_higgs}
\end{figure}

\begin{figure}[!htbp]
    \centering
      \subfloat[]{
        \includegraphics[width=0.45\textwidth]{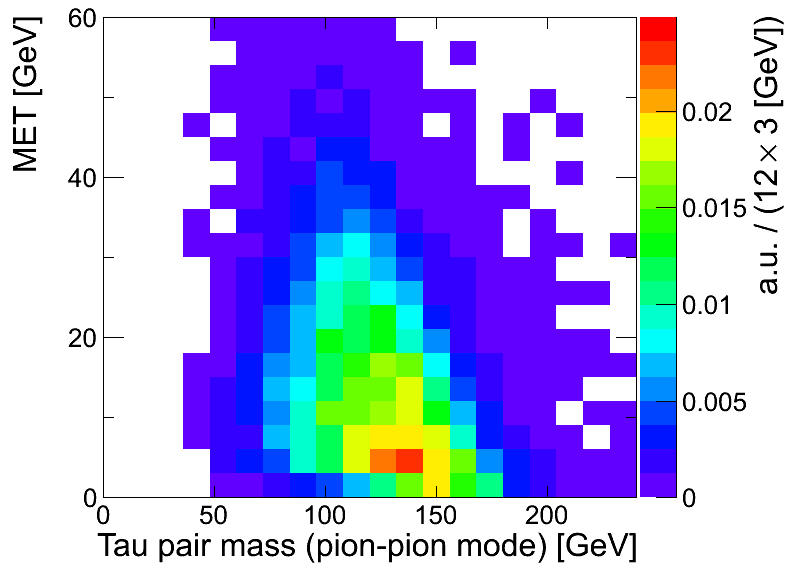}
\label{fig:msmr_pp_met_higgs}
      }
      \subfloat[]{
        \includegraphics[width=0.45\textwidth]{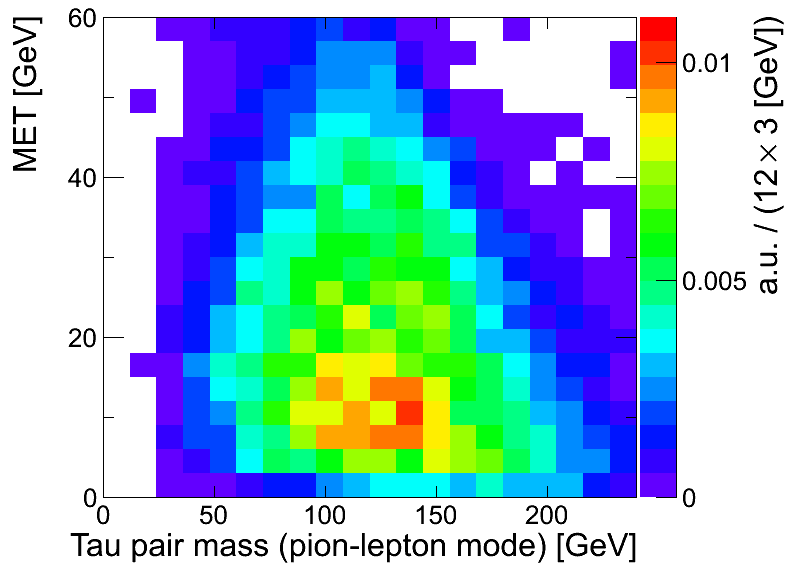}
\label{fig:msmr_pl_met_higgs}
      }
\hfill
      \subfloat[]{
        \includegraphics[width=0.45\textwidth]{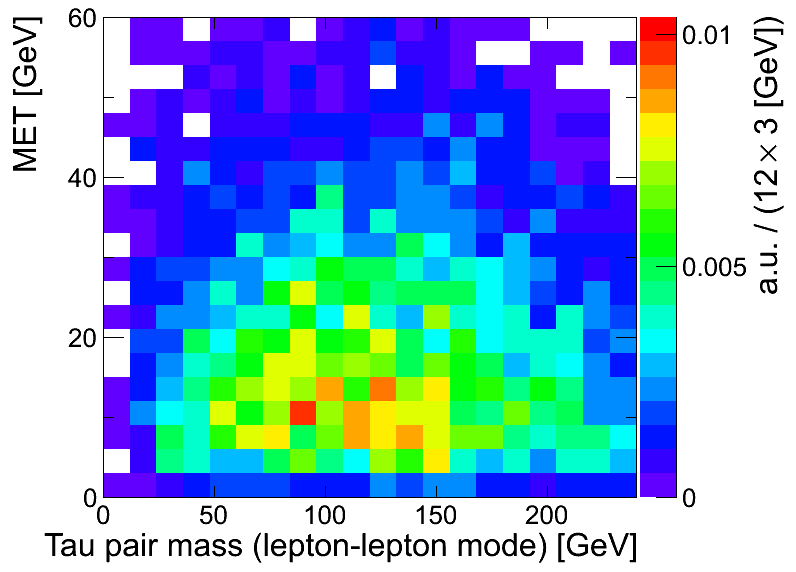}
\label{fig:msmr_ll_met_higgs}
      }
\caption{Correlation between stochastically reconstructed mass and $\MET$ in the $h \rightarrow \tau \tau$ process.}
\label{fig:msmr_pp_2d_higgs}
\end{figure}

The same procedure is repeated for $h \rightarrow \tau \tau$ processes, which comprise mostly the gluon-gluon fusion processes. 
I generate 100,000 events with the Higgs boson mass of 125 GeV.
As stated in sec.~\ref{sec:smr}, the stochastic mass-reconstruction technique is independent of  masses of heavy resonant particles.
Therefore, the peak structure naturally emerges at the right place ($ie$ the peak position shifts from the $Z$ boson to Higgs boson mass pole with respect to the previous sub-section).
The reconstructed masses are shown in the three categories in fig.~\ref{fig:msmr_higgs}, and its breakdown in pion multiplicities is shown in fig.~\ref{fig:msmr_npion_higgs}.
Again no strong linear correlation between the mass and $\MET$ is observed (see fig.~\ref{fig:msmr_pp_2d_higgs}).

\begin{figure}[!htbp]
    \centering
      \subfloat[]{
        \includegraphics[width=0.45\textwidth]{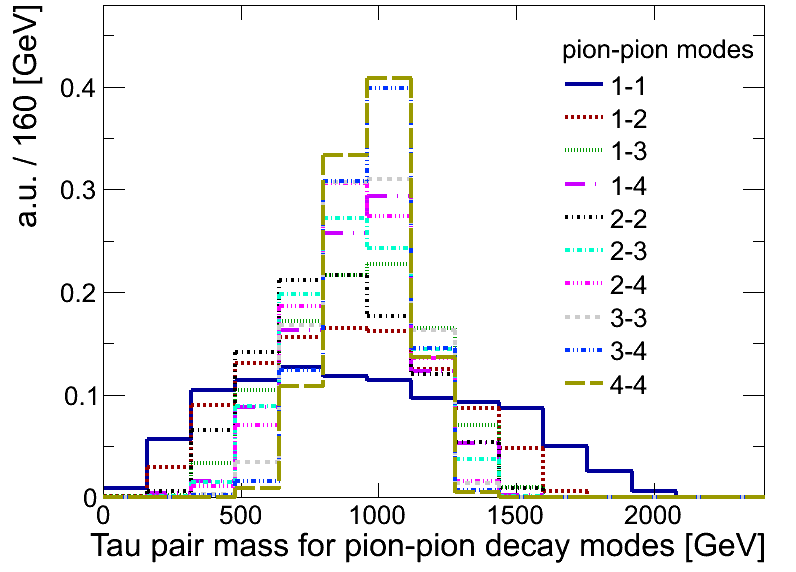}
\label{fig:msmr_pp_all_ZP}
      }
      \subfloat[]{
        \includegraphics[width=0.45\textwidth]{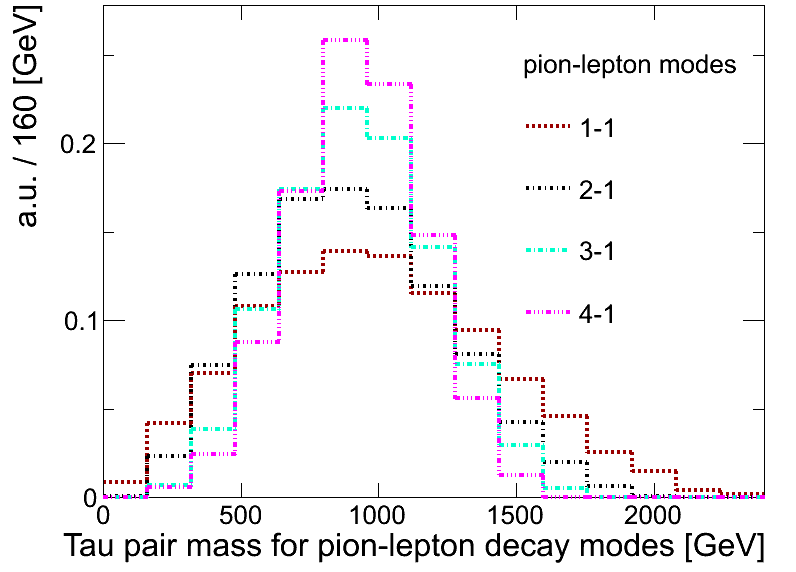}
\label{fig:msmr_pl_all_ZP}
      }
\caption{The stochastically reconstructed mass  for different combination of pion multiplicities in the $Z' \rightarrow \tau \tau$ process.}
\label{fig:msmr_npion_ZP}
\end{figure}

Similarly the same technique can be used for $Z' \rightarrow \tau \tau$ searches. 
As an example, I show stochastically reconstructed mass of 1 TeV $Z'$ bosons (see fig.~\ref{fig:msmr_npion_ZP}).

\section{QCD di-jet background}
\label{sec:qcd}
\begin{figure}[!htbp]
    \centering
      \subfloat[]{
        \includegraphics[width=0.47\textwidth]{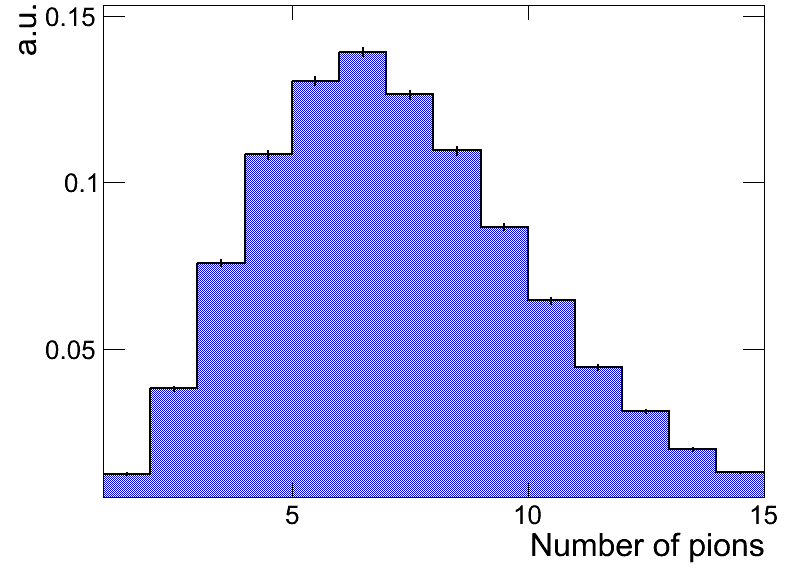}
 \label{fig:cdau1_qcd}
      }
      \subfloat[]{
        \includegraphics[width=0.47\textwidth]{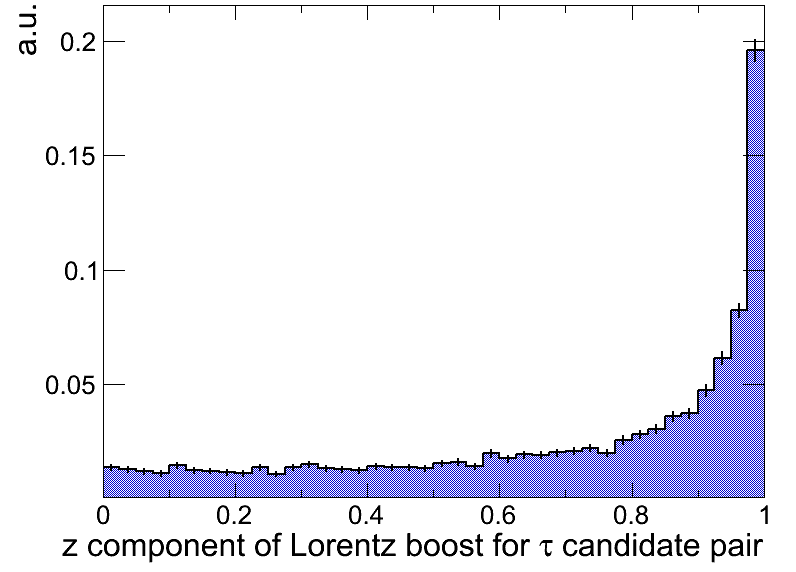}
\label{fig:bvis_qcd}
      }
\caption{The pion multiplicity and $z$-component of Lorentz boost vector for tau candidate pair distributions in the QCD di-jet process.}
\label{fig:dist_qcd}
\end{figure}

Background contribution from QCD di-jet events is relevant only in the final state consisting two hadronically decaying tau leptons, when jets are mis-identified as tau leptons.
In this section, I present a simple study for this background process.
The selection criteria is similar to the condition described in sec.~\ref{sec:z2tautau}, without matching jet constituents to tau daughters.
First I generate 100,000 QCD di-jet events (with $\hat{p_{T}}>40$ GeV), and then run the jet clustering algorithm with a distance parameter of 0.2, to define narrow jets (similar to hadronically decaying tau leptons). 
If a di-jet system satisfies the $\Delta \phi > \pi \times 0.9$ requirement and at least one pion is found in the jets, the event is kept for further studies.
I treat a set of pions inside the narrow jet as a tau candidate here.
The pion multiplicity and the $z$-component of Lorentz boost vector for the visible tau candidate pair are plotted in fig.~\ref{fig:dist_qcd}.
I keep jets with pion multiplicity up to five, motivated by the tau daughter multiplicity distribution (see fig.~\ref{fig:cdau1}).

\begin{figure}[!htbp]
    \centering
      \subfloat[]{
        \includegraphics[width=0.45\textwidth]{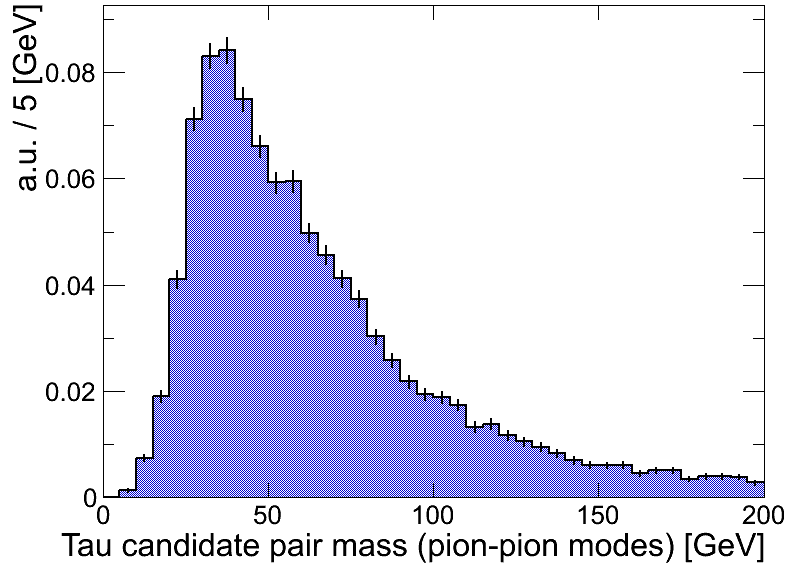}
\label{fig:msmr_pp_qcd}
      }
      \subfloat[]{
        \includegraphics[width=0.47\textwidth]{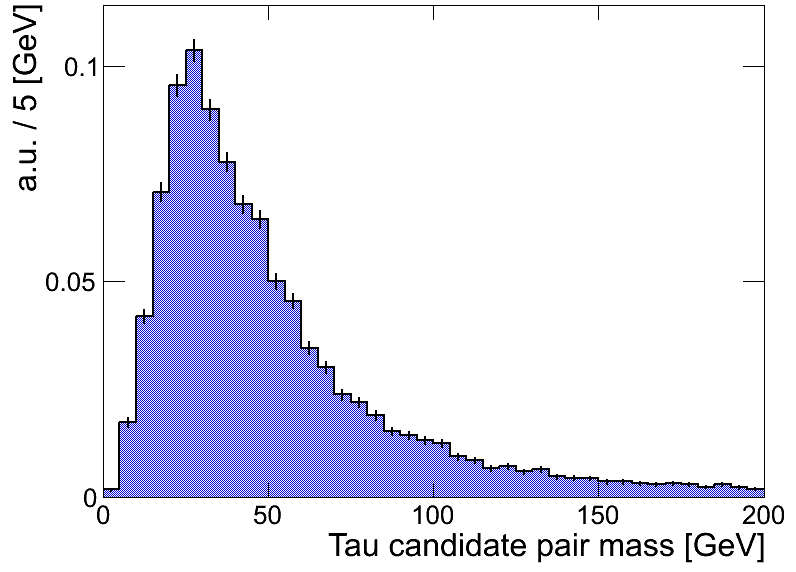}
 \label{fig:mzvis_qcd}
      }
\caption{The stochastically reconstructed mass  and visible tau candidate pair mass distributions in the QCD di-jet process.}
\label{fig:msmr_qcd}
\end{figure}

Mass reconstruction is performed by assuming $\tau \rightarrow \nu_{\tau}$ plus pion(s) decay modes ($ie$ the number of daughters = the number of pions plus one).
The stochastically reconstructed mass and visible tau candidate pair mass are plotted in fig.~\ref{fig:msmr_qcd}.
The distributions are monotonically falling in high-tail, and have an arbitrary cut off in low-tail as expected.

\section{Notes on usage}
\label{sec:note}
Actual implementation details of tau lepton identification algorithms vary among different experiments. 
In this section, I outline validation steps, although most of the procedure described in sec.~\ref{sec:smr} remains the same.
\begin{itemize}
\item first, with a SM boson ($eg$, a $Z$ or $h \rightarrow \tau \tau$) simulation sample, run tau identification algorithm and check how often lepton and pion decay modes are correctly reconstructed.
\item second, with the same simulated sample, check momentum and $p_{T}$ distribution of tau daughters. Typically there are minimum kinematic requirements on constituents of hadronically decaying tau leptons. When the requirements are too tight, then the expectation value would shift by a certain amount, and it might need a correction to account for the effect.
\item third, evaluate $\Delta \phi$ selection for the simulated samples with other dominant background processes. The $\Delta \phi > \pi \times 0.9$ selection is not necessarily optimal for a given signal of experimentalists' interests. The $\Delta \phi$ selection value should not be too small with respect to $\pi$, as it contributes to the width of reconstructed mass. 
\end{itemize}

The second point can be addressed by modifying the integration, for 4-body decay, for example, 
\begin{align}
f(x)_{4\text{-body}}^{'} & \propto \int_{\delta}^{1-x} (1-x-y)dy \\
& =  (y\times(1-x) - 1/2 \times y^2) |_{\delta}^{1-x} \\
& =  ( (1-x)^2 - 1/2 \times (1-x)^2) - (\delta \times (1-x) -1/2 \times \delta^{2})\\
& \approx 1/2\times(1-x)^2 - \delta \times (1-x)\\
f(x)_{4\text{-body}}^{'} & \approx \left ( \frac{3}{1-6\delta} \right ) (1-x) (1-x-2\delta)  \\
{\langle x_{4\text{-body}} \rangle }^{'} & \approx (1/4) \left ( \frac{1-4\delta}{1-6\delta} \right )
\end{align}
where $\delta$ is a small selection bias $\approx x_{\text{cut}} / x_{\text{mom}} \ll 1$, and the lowest order $\delta$ terms are kept.
With 1 GeV threshold on daughter momenta for a 90 GeV boson, the shift in the mean momentum is $\approx$ 5\% with respect to zero threshold.
If different sets of selection are used for neutral pions and charged pions, the mean value may be estimated with simulated samples.
Evaluation of the shifted mean is straight forward; plotting stochastically reconstructed mass, and calculating its mean value for a given tau decay mode.
The calculation can be repeated for all decay modes of interests to fine tune the expectation value matched to the pole mass in the simulated samples. 

For a small $n$, the bias induced by the kinematic requirements on tau candidates can be large as seen in the 1-1 pion mode (see fig.~\ref{fig:msmr_pp_all}).
If tau candidates ($ie$ visible part of tau leptons) need to satisfy jet $p_{T}$ above a certain threshold, this requirement can be entirely on one pion.
The corresponding momentum space is reduced by the threshold; $ie$ from 1 to  $\approx 1-\epsilon $ where $\epsilon$ is a ratio of averaged momentum threshold induced by the kinematic selection to mother momentum.
For example, the 10 GeV selection is not small compared to 45 GeV, if the entire 10 GeV needs to be accounted for by a single pion.
This burden is shared by multiple particles for larger $n$, and therefore this bias is less significant for larger $n$ decay modes.
The weights applied to pions in small $n$ modes can be adjusted with simulated samples when mother particle masses are relevant to the selection thresholds.

The third point is required by construction; I select events that are not significantly Lorentz boosted in either the $x$- or $y$-direction to make a connection between a tau pair.
In general an invariant mass of a tau pair is not $m=2\sqrt{E_{\tau1}E_{\tau2}}$ but $m=\sqrt{2E_{\tau1}E_{\tau2}\gamma^2(1-\beta^2cos\theta_{a}^2)(1-cos\theta_{b})}$ where $E_{\tau}$ is a tau energy; $\gamma$, $\beta$, and $\theta_{a}$ connect the rest frame of the mother particle to the inertial reference frame where an angle $\theta_{b}$ between a tau pair is measured.
Negligible Lorentz boost in the transverse direction implies that $\gamma^2(1-\beta^2cos\theta_{a}^2) \rightarrow 1$, and $(1-cos\theta_{b}) \rightarrow 2$, by Lorentz boosting along the -$z$-direction.
When this condition is no longer met, the collinear approximation works well with measured $\MET$ values.

In this first paper on the technique, I have restricted my interests to tau pairs originating from heavy mother particles, and events where the collinear approximation technique degrades. 
It can also be used to estimate momentum of each tau lepton separately for large $n$ as well as for processes yielding only one tau lepton: $eg$ $W \rightarrow \tau \nu$ process. 
However, it is necessarily to incorporate a $\MET$ value to reconstruct a transverse mass or $m_{T2}$~\cite{mt2} of $W$-like bosons in such a case because the trajectory of the neutrino is unknown. 
The one-tau case is beyond the scope of this paper, and left for future studies.

\section{Summary}
\label{sec:summary}
The stochastic mass-reconstruction is a novel technique to estimate resonant masses of heavy particles decaying to tau lepton pairs.
It is simple and straightforward to implement thanks to its analytic form and summation of daughter momenta.
It is a new variable which can be used in conventional di-tau analyses that utilize another kinematic variable such as a \MET, because the new technique does not explicitly rely on other kinematic variables but the properties of a pair of tau leptons.
Since the peak structure associated with mother particle masses naturally appears, the technique can be used for $Z$ and SM Higgs boson measurements, as well as exotic particle searches such as a $Z'$ boson decaying to tau leptons.

\section*{Acknowledgments}
\label{sec:ack}
I would like to thank Maxwell Chertok (University of California at Davis) and Aron Soha (Fermilab) for fruitful discussions and constructive suggestions. 
Fermilab is operated by Fermi Research Alliance, LLC under Contract No. De-AC02-07CH11359 with the United States Department of Energy.


\begin{thebibliography}{99}
\bibliographystyle{unsrt}

\bibitem{atlasDiscH}
ATLAS collaboration, 
\textit{Observation of a new particle in the search for the Standard Model Higgs boson with the ATLAS detector at the LHC}, 
Phys. Lett. B 716 (2012) 1, doi:10.1016/j.physletb.2012.08.020, arXiv:1207.7214

\bibitem{cmsDiscH}
CMS collaboration, 
\textit{Observation of a new boson at a mass of 125 GeV with the CMS experiment at the LHC}, 
Phys. Lett. B 716 (2012) 30, doi:10.1016/j.physletb.2012.08.021, arXiv:1207.7235

\bibitem{ca}
R. K. Ellis  $et \, al.,$
\textit{Higgs decay to $\tau^{+} \tau^{-}$: possible signature of intermediate mass Higgs bosons at high energy hadron colliders}
Nucl. Phys. B 297 (1988) 221

\bibitem{mmc}
A. Elagin $et \, al.$,
\textit{A New Mass Reconstruction Technique for Resonances Decaying to di-tau},
Nucl. Instr. Meth. Phys. Res. 654 (2011) 481, doi:10.1016/j.nima.2011.07.009, arXiv:1012.4686v2

\bibitem{cmsTau}
CMS Collaboration, 
\textit{Performance of $\tau$-lepton reconstruction and identification in CMS},
 J. Instrum. 7 (2012)1001, doi:10.1088/1748-0221/7/01/P01001, arXiv:1109.6034 

\bibitem{pdg}
K.A. Olive $et \, al.$ (Particle Data Group), 
\textit{The Review of Particle Physics},
Chin. Phys. C 38 (2014) 090001 and 2015 update

\bibitem{atlasHiggs}
ATLAS collaboration,
\textit{Evidence for the Higgs-boson Yukawa coupling to tau leptons with the ATLAS detector},
JHEP 04 (2015) 117, doi:10.1007/JHEP04(2015)117, arXiv:1501.04943

\bibitem{cmsHiggs}
CMS Collaboration, 
\textit{Evidence for the 125 GeV Higgs boson decaying to a pair of $\tau$ leptons},
JHEP 05 (2014) 104, doi:10.1007/JHEP05(2014)104, arXiv:1401.5041

\bibitem{bell}
J.S. Bell and R. Jackiw, 
\textit{A PCAC Puzzle: $\pi^{0} \rightarrow \gamma\gamma$ in the $\sigma$-model},
Nuovo Cim. A 51 (1969) 47

\bibitem{adler}
S.L. Adler, 
\textit{Axial vector vertex in spinor electrodynamics},
Phys. Rev. 177 (1969) 242

\bibitem{pythia}
T. Sj$\ddot o$strand $et \, al.,$
\textit{An Introduction to PYTHIA 8.2},
Comput. Phys.Commun. 191 (2015) 159, doi:10.1016/j.cpc.2015.01.024, arXiv:1410.3012

\bibitem{superRazor}
M.R. Buckley $et \, al.,$
\textit{Super-razor and searches for sleptons and charginos at the LHC},
Phys. Rev. Lett.  89 (2014) 055020, doi:10.1103/PhysRevD.89.055020, arXiv:1310.4827

\bibitem{antikt}
M. Cacciari $et \, al.,$ 
\textit{The anti-$k_t$ jet clustering algorithm},
JHEP 04 (2008) 63, doi:10.1088/1126-6708/2008/04/063, arXiv:0802.1189

\bibitem{atlasTau}
ATLAS Collaboration,
\textit{Identification and energy calibration of hadronically decaying tau leptons with the ATLAS experiment in pp collisions at $\sqrt{s}$ = 8 TeV},
Eur. Phys. J. C75 (2015) 303, doi:10.1140/epjc/s10052-015-3500-z, arXiv:1412.7086

\bibitem{mt2}
C.G. Lester $et \, al.,$
\textit{Measuring masses of semi-invisibly decaying particles pair produced at hadron colliders},
Phys. Lett. B 463 (1999) 99, doi:10.1016/S0370-2693(99)00945-4, arXiv:hep-ph/9906349


\end{thebibliography}
\end{document}